\documentclass{PoS}

\title{Hints of High Core Faraday Rotations from a Joint Analysis of 
VLBA and Optical Polarization Data}

\ShortTitle{Hints of High Core Faraday Rotations}

\author{\speaker{Juan Carlos Algaba}\\
        University College Cork\\
        E-mail: \email{algaba@physics.ucc.ie}}

\author{Denise C. Gabuzda\\
        University College Cork\\
        E-mail: \email{gabuzda@physics.ucc.ie}}

\author{Paul S. Smith\\
        Steward Observatory, University of Arizona\\
        E-mail: \email{psmith@as.arizona.edu}}

\abstract{Although the continua of radio-loud Active Galactic Nuclei (AGN) 
are typically dominated by 
synchrotron radiation over virtually the entire spectrum, it is not clear 
whether the radio and higher-frequency emission originate in the same or 
different parts of the jet. Several different radio--optical correlations based 
on polarization data have been found recently, suggesting that the optical 
and radio polarization may be closely related, and that the corresponding 
emission regions may be cospatial\cite{Gabuzda06,Jorstad07,DArcangelo07}. Our 
joint analysis of optical and VLBA polarization data for a sample of about 
40 AGNs shows that, after correction for the 
inferred VLBA core Faraday rotations, most BL Lac objects and some quasars 
have aligned VLBA-core and optical polarizations, although many quasars 
also show no obvious relationship between their VLBA-core and optical 
polarization angles. This may indicate that not all AGNs have cospatial regions 
of optical and radio emission in their jets. However, another possibility is 
that some of the 7mm--2cm VLBA cores have Faraday rotations of the order of 
several tens of thousand of rad/m$^2$,  which were not properly fit by our 
three-frequency data due to $n\pi$ ambiguities in the observed polarization 
angles, leading to incorrect subtraction of the effects of the core 
Faraday rotation, and so incorrect ``zero-wavelength'' radio polarization 
angles. The possibility of such high core Faraday rotations is supported by 
the results of the parsec-scale Faraday-rotation studies of 
Zavala \& Taylor\cite{ZandT04} and Jorstad et al.\cite{Jorstad07}.}

\FullConference{The 9th European VLBI Network Symposium on The role of VLBI
in the Golden Age for Radio Astronomy and EVN Users Meeting\\
		 September 23-26, 2008\\
		 Bologna, Italy}

\begin{document}

\section{Introduction}
The term Active Galactic Nucleus (AGN) refers to the existence of very 
energetic phenomena occurring in the centers of some galaxies that cannot be 
attributed to stars. In the standard model, AGNs consist of a black hole 
surrounded by an accretion disk. There is good evidence that relativistic jets 
in AGNs are powered by nonthermal synchrotron radiation; i.e., a flow of 
radiating, accelerated electrons embedded in a magnetic field, which dominates 
the continua of electromagnetic radiation over a wide window, extending from 
radio to soft X-ray frequencies.

It is natural to suppose that all these wavebands should
share common properties, and that variability should be correlated
across the spectrum. However, consideration of the physics of AGN
jets leaves it unclear whether this will, in fact, be the case. In some 
models \cite{MarscherGear85}, higher-energy emission is generated closer 
to the base of the jet, so that emission at different wavelengths arises 
in different regions of the jet with different physical properties (e.g. 
different magnetic-field configurations), while, in
others, the high-energy and radio emission can be cospatial, at least for certain
combinations of jet geometry and particle flow \cite{Ghisellini85}. In addition,
even if emission from various wavebands are intrinsically closely
related,  there are a variety of wavelength-dependent extrinsic processes, 
such as Faraday Rotation and scintillation, that can give rise to different 
observed properties at different wavelengths. Thus, it is important to
correct for these extrinsic effects if we wish to test for correlated behaviour
in different wavebands.

There is recent evidence of a much closer link between the radio and 
high-energy synchrotron radiation that was originally indicated by previous studies.
For example, \cite{Jorstad01} found a
tendency for gamma-ray flares to occur several months after the 
births of new VLBI component, suggesting these flares occur in these 
radio components, at appreciable distances down the jet. 
Polarization information has played a key role in revealing evidence for
other correlations: \cite{Gabuzda06} 
demonstrated a strong tendency for the simultaneously measured optical and 
Faraday-rotation-corrected VLBI core polarization angles to be aligned in about 
a dozen BL Lacs, and \cite{Jorstad07} observed similar behaviour 
for the optical and high-frequency radio polarization angles in a sample of 
highly polarized AGNs. Also, \cite{DArcangelo07} observed a rapid, 
simultaneous rotation in the optical and 7mm VLBA-core polarization angles
in 0420-014. All these results support the idea that the radio to optical, 
and even gamma-ray, emission may be more closely related than was previously 
thought. 

\section{Observations and Results}
With the aim of verifying the results of \cite{Gabuzda06} and searching for 
further evidence for optical--VLBI polarization correlations, we observed 
an additional $\sim 30$ AGNs, including BL Lac objects 
and both high- and low-polarization quasars (here, the degree
of polarization refers to the optical waveband \cite{Moore84}),
thus providing us with a sample of 40 AGNs for our analysis. We 
obtained 7mm+1.3cm+2cm VLBA polarization data and nearly simultaneous optical 
polarization data with the Steward Observatory 2.3m telescope in three 
24-hour sessions, on November 1, 2004, March 16, 2005 and September 26, 2005. 
The data reduction and imaging for the radio data were done with the NRAO 
Astronomical Image Processing System (AIPS) using standard techniques (see, 
e.g. \cite{Gabuzda06}).

The optical polarization observations spanned the VLBA observation runs 
(October 30--November 2, 2004; March 15--17, 2005; and September 25--29, 2005).
These data were acquired using the SPOL imaging/spectropolarimeter
\cite{Schmidt92}. On various nights, the instrument was configured for either 
imaging polarimetry using a KPNO ``nearly mould'' R filter (6000--7000\AA) 
or spectropolarimetry using a 600 line/mm diffraction grating. The data 
acquisition and reduction closely followed those described in\cite{Smith07}. The 
spectropolarimetric observations were averaged over the R-filter bandpass for 
direct comparison to the imaging polarimetry. 

Since we have VLBA polarization data at three wavelengths, we expected to be 
able to correct for Faraday rotation in the region of the compact core, 
enabling a comparison between the ``zero-wavelength'' radio-core and optical 
polarization angles. If the optical and radio emission is cospatial, and the
emitting region is optically thin at both  
wavebands, the difference $\Delta\chi$ between the optical $\chi_{opt}$ and 
Faraday-rotation-free radio-core $\chi_0$ polarization angles should 
be close to zero. On the other hand, if
the VLBI-core emission is optically thick at the observed radio
wavelengths, this should give rise to an offset of $\Delta\chi=90^{\circ}$.
Thus, our ``null hypothesis'' was that the distribution of $\Delta\chi$
values would be dominated by two peaks: one near $0^{\circ}$ and one near
$90^{\circ}$. However, this was not the case for the new observations: although 
the complete sample of BL Lac objects displayed
an overall clear peak near $0^{\circ}$,
the quasars display a flatter distribution, possibly with a weak
peak around $\sim50^{\circ}$\cite{Algaba08}. We discuss in section 3 the possible
origins of this difference in behaviour shown by the BL Lac objects and the
quasars in our sample.

\section{Discussion}
One obvious possibility is that there is indeed no correlation between the
optical and VLBI-core polarization angles for both the high-polarization and
low-polarization quasars. In other words, these results may provide evidence
that the optical and VLBI-core emission in quasars is usually not cospatial, 
with the different emitting regions having different properties. If so, this 
would seem to indicate a difference between the geometries or physical
conditions of the jets of quasars and BL Lac objects that is worthy of 
further study.

However, the hint of a possible weak peak in the $\Delta\chi$ distribution 
for the quasars suggests another possibility: that the VLBI cores are
subject to internal Faraday rotation. In this case, the rotation of the radio
polarization angle 
initially obeys the $\lambda^2$ law characteristic of
external Faraday rotation, but then 
saturates at rotations of about 
$40-50^{\circ}$ with increase in the observing wavelength. If all
three of our wavelengths were in this regime, this could appear
as a small core Faraday rotation, leading to an inferred $\chi_{0}$ of
about $40-50^{\circ}$ away from its true value. Our past investigation
showed no evidence for
enhanced depolarization in the quasar cores compared to the BL Lac cores,
leading us to discard this idea as an explanation for the behaviour of the
quasars in our sample as a whole \cite{Algaba08}. 

There is, however, another possibility: that the 7mm--2cm quasar cores are 
subject to appreciably higher local but external Faraday rotations than the 
BL Lac cores at these wavelengths. This would be consistent with the tendency
for quasar cores to have higher Faraday rotation measures than BL Lac cores
at 2cm--4cm \cite{ZandT04}. In fact, core rotation
measures as high as tens of thousands of rad/m$^2$ have been reported previously 
\cite{Jorstad07,ZandT04}. In the
case of high core rotation measures, we encounter problems with 
$n\pi$ ambiguities when fitting the rotation measures, leading to incorrect
determinations of both the rotation measure and $\chi_0$.

Our data suggest that such ambiguities may be 
present. For example, if we observe a roughly $90^{\circ}$ change in the observed 
polarization angles between two radio wavelengths, it is natural to 
suppose that this is due to a transition between the optically thick and 
optically thin regimes. This can be verified by
examining the observed spectral indices and degrees of polarization of the 
VLBI core. However, this supporting evidence is not found in most cases where we 
observe polarization-angle rotations by roughly $90^{\circ}$ between neighbouring
wavelengths, suggesting that for these objects, high core Faraday 
rotations may be relevant. In addition, some of the
core rotation measures indicated by our observations
are low compared to the typical values deduced by \cite{ZandT04}, 
despite taking
into account the fact that our 7mm--2cm observations probe somewhat smaller 
scales, where we expect both the electron density and magnetic-field strength
to be higher. 

\begin{figure}
\begin{center}
\includegraphics[width=.80\textwidth]{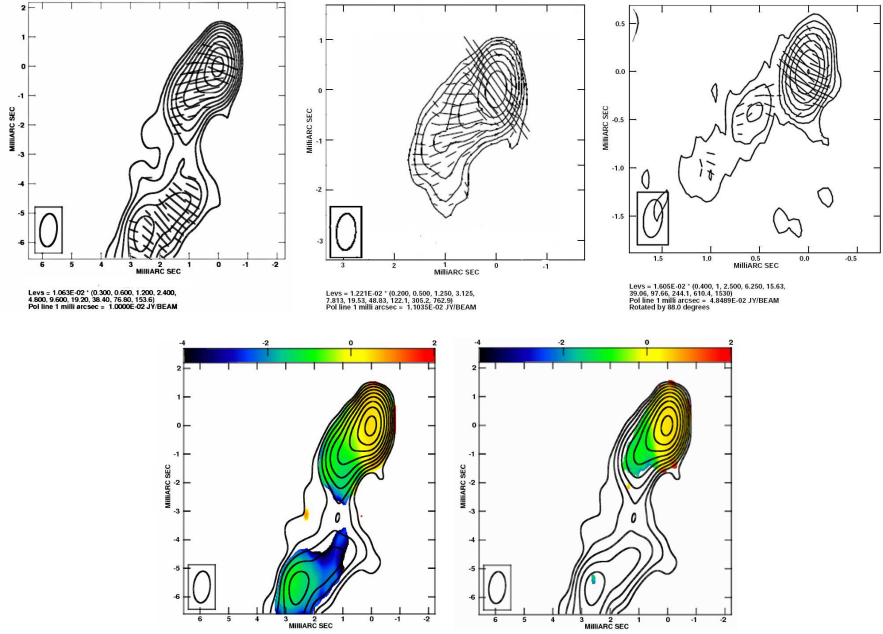}
\caption{Top: total intensity maps with superposed polarization vectors
for 2230+114 at 2cm (left), 1.3cm (middle) 
and 7mm (right). Bottom: 2cm--1.3cm (left) and 1.3cm--7mm (right) spectral-index 
distributions for this source; both core spectral indices are $\sim0.46$, giving
no evidence for a change in optical depth in the observed frequency range.}
\end{center}
\vspace{-0.8cm}
\end{figure}
\begin{figure}
\begin{center}
\includegraphics[width=.70\textwidth]{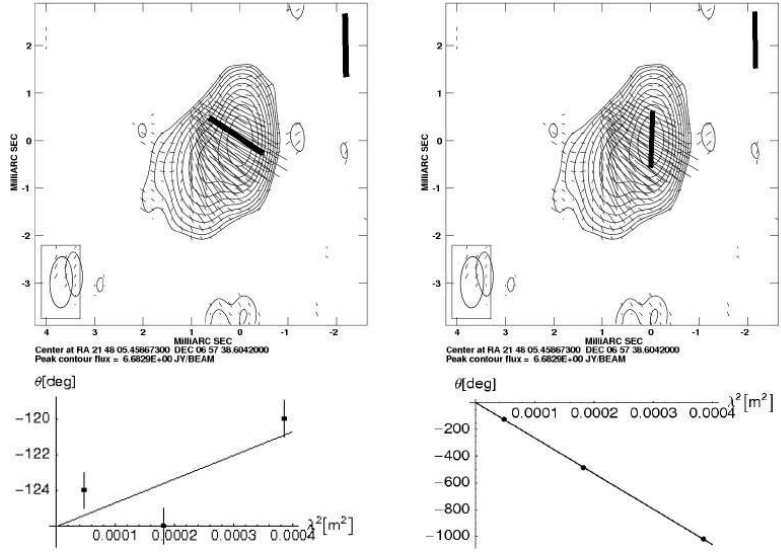}
\caption{Two alternative rotation-measure fits for 2145+067. Left: the
rotation-measure fit obtained for the nominal observed radio polarization
angles. Right: another possible fit allowing for possible $n\pi$ ambiguities
in the observed radio polarization angles. The thick lines in the corner and 
in the core show the optical and VLBI-core Faraday-corrected polarization
angles, respectively.}
\end{center}
\vspace{-0.8cm}
\end{figure}

Two examples are shown in Figs. 1 and 2. Fig. 1 shows our images for the quasar 
2230+114. The nominal observed
values for the polarization angles are $70^{\circ}$ at 2cm,
$34^{\circ}$ at 1.3cm and $94^{\circ}$ at 7mm, which clearly do not yield a 
reasonable $\lambda^2$ fit. A good fit is obtained if both the 2cm and 1.3cm
polarization angles are rotated by $90^{\circ}$, with the implied offset 
between $\chi_0 - \chi_{opt} \simeq 60^{\circ}$. An inspection 
of the core degrees of polarization at our three wavelengths and the 2cm--1.3cm 
and 1.3cm--7mm core spectral indices ($\alpha\simeq 0.46$) provide no
evidence for an optical-depth transition in our wavelength range, but 
suggest that the VLBA core is probably optically thick 
at all three wavelengths. Alternatively, a comparably good fit is
obtained if the 1.3cm polarization angle is rotated by $\pi$ and the 2cm
polarization angle by $2\pi$; in this case, the inferred rotation measure is
about $16,700$~rad/m$^2$ and the inferred zero-wavelength polarization
angle is $\chi_0\simeq 45^{\circ}$, offset from $\chi_{opt}$ by roughly
$80^{\circ}$, close to the offset expected if the radio and
optical emission regions are cospatial, but the radio core emission is optically
thick.

Similarly, Fig. 2 shows two possible rotation-measure fits for 2145+067. 
The left option shows the fit to the nominal observed radio-core polarization
angles, which agree with each other to within about $6^{\circ}$. This
fit is acceptable, and implies an offset $\Delta\chi\simeq 
53^{\circ}$. However, if we add $2\pi$ to the 1.3cm and $5\pi$ to the 
2cm polarization angles (right panels in Fig.~2), we obtain a comparably good
fit that corresponds to a much larger but still reasonable core rotation 
measure of about +47,000 rad/m$^2$ and a zero-wavelength polarization angle 
of about $-74^{\circ}$. This fit implies a near alignment between the optical and 
VLBI-core polarization angles, $\Delta\chi\simeq 5^{\circ}$, which is consistent
with an optically thin VLBI core
($\alpha = -0.30$ between 2cm and 1.3cm and $\alpha = -0.60$ between
1.3cm and 7mm).

We can find at least one acceptable fit with a high core rotation measure for 
essentially every source in our sample, usually implying rotation measures of
the order of several tens of thousands of rad/m$^2$. In a few cases, we can
identify an unambiguous best fit, but in others two or more 
different possible core rotation measures are plausible.
Some of these fits imply intrinsic 
radio core polarization angles that are either aligned with or orthogonal
to the optical polarization angle, consistent with the observed core spectral
indices.

It can be argued that indiscriminately adding or subtracting some 
number of $\pi$ rotations to the observed VLBA polarization 
angles, we will eventually obtain a good fit,
even if the inferred rotation measure has no physical basis.
For this reason, we held the 7mm polarization angle fixed at its
observed value, and considered rotations of no more than $\pm 5\pi$ for the 2cm
polarization angles. The corresponding rotation measures range up to roughly
$\pm50,000$~rad/m$^2$, and are plausible, since we are probing regions of
higher electron density and stronger magnetic fields. Similar rotation measures
were found previously \cite{Jorstad07,ZandT04}. In addition, allowing
for a high core rotation measure 
yields a good fit to the rotated polarization angles, there is no guarantee
that the corresponding zero-wavelength core polarization angle will show some
correlation with the optical polarization angle.

Therefore, it is likely that some of our alternative high core rotation 
meausures are correct, but they must be confirmed. It remains possible that
internal Faraday rotation is influencing the
observed polarization angles in some cores. Unfortunately, it is impossible
to distinguish between the various scenarios with our current 
three-wavelength data. For this reason, we have obtained new VLBA polarization
observations of 8 sources from our sample at 5 wavelengths from 7mm to 2cm. 
The new data will improve
our ability to identify high external Faraday rotation and the signature of 
internal Faraday rotation if present in the cores of these AGNs. Depending on
these results, we may propose analogous observations for a larger sample.

The possible presence of high core rotation measures has certain
interesting implications. First, it would suggest that the physical 
conditions in the sub-parsec-scale jets in quasars are more 
extreme than has previously been thought. Second, a friendly warning for 
observers: we must be careful when observing polarization in quasar cores. 
The possible presence of high core rotation measures should be taken into
consideration when planning polarization observations. For example, it is
clear from our results that three wavelengths may not be sufficient to
unambiguously derive reliable Faraday rotations, even at short VLBI 
wavelengths. In addition, it seems likely that we should not assume that
the 7mm polarization angles are a good aproximation to the intrinsic 
polarization, as has usually been assumed. Furthermore,
many of the MOJAVE\cite{MOJAVE} 2cm core polarization angles are likely
subject to appreciable Faraday rotation.

\end{document}